\documentclass[twocolumn,amsmath,ammsymb]{revtex4-1}
\usepackage{epsfig}
\usepackage{amsmath}
\usepackage{amssymb}
\usepackage{graphicx}
\usepackage{dcolumn}
\usepackage{soul}
\usepackage{bm}
\usepackage[usenames]{color}

\begin{document} 

\preprint{}

\title{Spontaneous symmetry breaking of an interacting Chern insulator on a topological square lattice}
\author{Yi-Xiang Wang}
\affiliation{School of Science, Jiangnan University, Wuxi 214122, China.}
\affiliation{Department of Physics and Astronomy, University of Pittsburgh, Pittsburgh, Pennsylvania 15260, USA}
\author{Dong-Xiang Qi}
\affiliation{National Laboratory of Solid State Microstructures, School of Physics, and Collaborative Innovation Center of Advanced Microstructures,
Nanjing University, Nanjing 210093, China.}

\date{\today}

\begin{abstract}
The interplay between topology and correlation lies at the forefront of the modern condensed matter physics.  In this work, we study the extended fermion-Hubbard model, including the onsite as well as the nearest-neighbor repulsive interactions, on a topological square lattice that supports the Chern insulator.  Within the mean-field method, we find that the spontaneous symmetry breaking (SSB) charge density wave or antiferromagnetic insulator dominates the system when the onsite or NN interactions are strong enough.  It is interesting that the antiferromagnetic Chern insulator will appear in the phase diagram when there is an explicitly nonvanishing sublattice potential.  In addition, we explore how a finite-size ribbon structure affects the phase diagram and point out that the critical interaction for SSB occurs with weaker strength than the bulk system.  
\end{abstract}

\maketitle

\section{Introduction}

Topological matter represents one of the most intriguing frameworks to realize unconventional physics \cite{M.Z.Hasan, X.L.Qi, C.X.Liu, C.L.Kane}.  After its first theoretical proposal by Haldane thirty years ago \cite{Haldane}, the Chern insulator (CI) has recently been successfully observed in magnetic-doped topological insulators \cite{C.Z.Chang} as well as in cold-atom experiments \cite{G.Jotzu} with shaking lattice technique.  As the topological bands of CI are described as the noninteracting fermion model, a crucial question arises that to what extent such topological bands are stable to interactions.  Or if they are unstable, what will happen otherwise?  

Many previous works explored these questions with the Haldane-Hubbard model.  The Hubbard model of spin$-\frac{1}{2}$ fermions describes on-site repulsive interactions and can lead to highly nontrivial correlation effects.  A commonly accepted viewpoint is that the repulsive interactions can drive the formation of unconventional phases in the topological system \cite{J.He1, J.He2, A.M.Cook, W.Zheng, T.I.Vanhala, V.S.Arun, D.Prychynenko, K.Jiang, J.Imriska}.  For example, it was found that besides the quantum Hall phase with Chern number $C=2$ and the band insulator (BI), the Haldane-Hubbard model can also accommodate the Mott insulating phase and quantum Hall phase with $C=1$ \cite{T.I.Vanhala,J.He1}.  More interestingly, the chiral noncoplanar magnetic orders are uncovered in the enlarged four-site, or even six-site unit cell of Haldane-Hubbard model \cite{V.S.Arun,W.Zheng}.  In Ref. \cite{Y.C.Zhang}, the attractive Haldane-Hubbard model was also studied and the topological superfluid with Chern number $C=2$ was revealed for intermediate attractive strength.  Motivated by these, here we focus on the extended Hubbard model, including the onsite repulsive interactions as well as the nearest-neighbor (NN) repulsive interactions, on a topological square lattice that can support the CI.  Compared with the honeycomb lattice, the square lattice is more feasible to be implemented in cold-atom systems \cite{M.Aidelsburger, H.Miyake, Z.Wu}.  

An important finding in our work is the appearance of the interaction-driven antiferromagnetic Chern insulator (AFCI).  The characteristic of the AFCI is that it incorporates the spontaneous symmetry breaking (SSB) long-range magnetic order as well as the nontrivial bands \cite{K.Jiang, Y.X.Wang2018}.  In previous works about the magnetic topological insulators (TIs) \cite{R.Mong, C.Fang}, the AFCI phase was suggested to exist in a TI thin film, with the antiferromagnetic (AFM) spin order being induced by the AFM substrate.  However, it requires the lattice matching or commensuration between the substrate and thin film, which makes it complicated for real electronic materials.  Here we suggest another feasible route to realize the AFCI through the correlation effect in a CI model based on a square lattice.

Another question is when the two-dimensional (2D) system owns the ribbon structure \cite{J.Cao}, how does the finite width affects the SSB and the phase transitions.  In the CI phase, the ribbon supports the edge states, which can cause the fermion density decreasing exponentially from the edge sites to the central ones \cite{K.Nakada, P.Delplace}.  The finite density of states of the edge states may make them susceptible to either charge or spin orderings, even for vanishingly small interactions.  Then what happens to the whole ribbon system will be explored here.

With the help of the mean-field (MF) theory, we solve the extended fermion-Hubbard model on a topological square lattice self-consistently.  The MF method is qualitatively reliable, as it captures the essential correlations with the change of parameters in a many-body system.  The main results are as follows:  (i) According to the static susceptibilities, we judge that among various long-range orders, only charge density wave (CDW) or AFM order may dominate the system.  (ii) We analyze the renormalized fermion mass and energy gap when CDW or AFM order is present.  In determining the fermion occupation number, the competition mechanism between the onsite and NN interactions is revealed.   (iii) We calculate the interacting-dressed bulk phase diagrams.  Due to the vanishing density of states around the Dirac points, the AFM and CDW  will set in for sufficiently strong on-site and NN repulsions, respectively.  Especially interesting is the emergence of the AFCI when the sublattice potential is explicitly nonvanishing.  The survival condition for the AFCI is analyzed that only when the time-reversal symmetry (TRS) is truly broken can such a novel phase appear.  (iv) In addition, we study the interacting square lattice ribbon structure.  The results show that in the ribbon system with finite width, through the proximity effect, the local edge orderings can induce the SSB long-range orders in the whole system at the weaker critical interaction strength.  Our work may be helpful in understanding the effect of short-ranged interactions in CI and may shed some lights in future topological electronic devices.

\section{Model}

\begin{figure}
	\includegraphics[width=6cm]{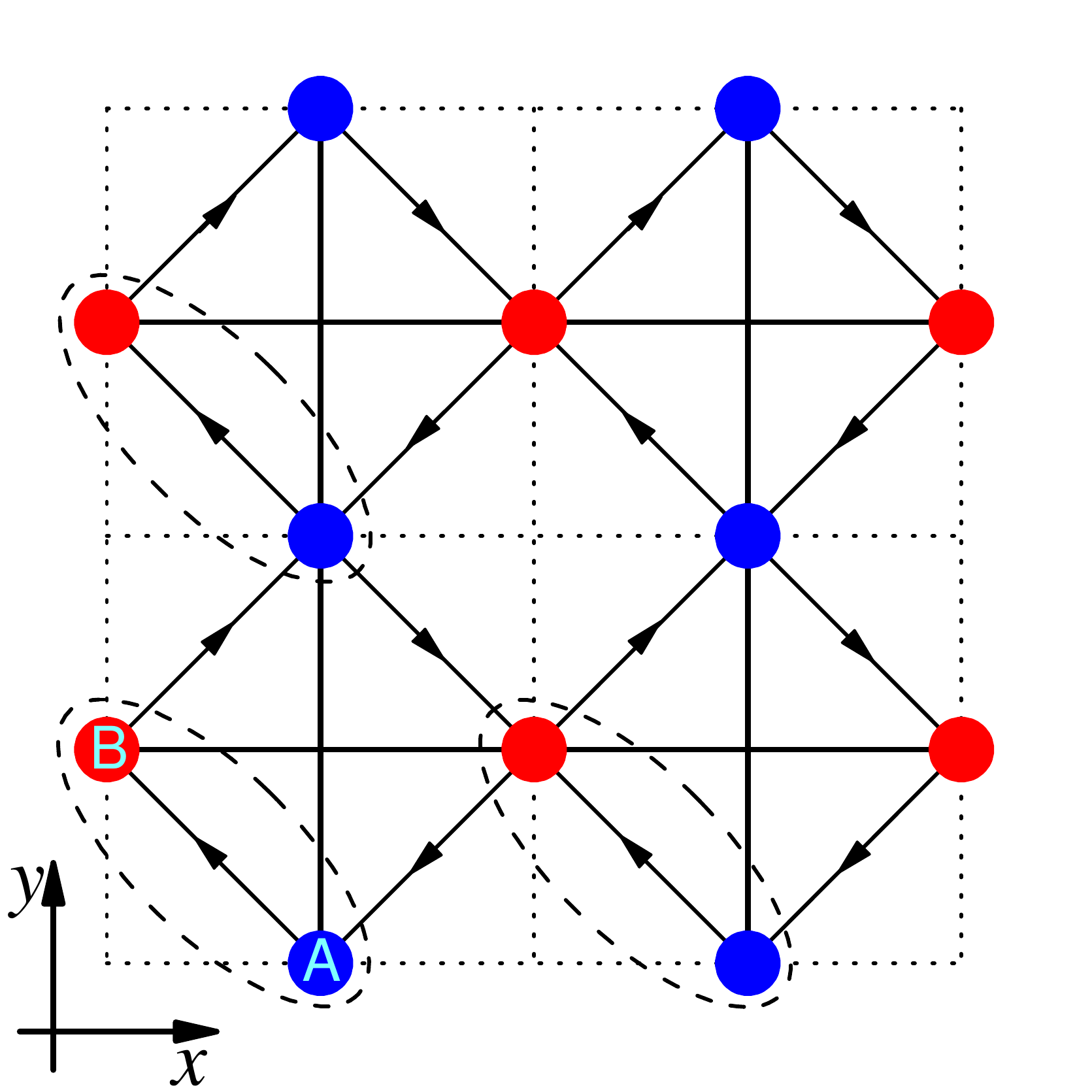}
	\caption{(Color online) Schematic plot of a topological square lattice, with the unit cell in the dashed oval including two sublattices A and B.  The arrows show the direction of positive phase winding for the complex NN hoppings, which are responsible for the topological properties of the model. The NNN hoppings are anisotropic in the perpendicular directions, $t_2$ and $-t_2$, indicated by the solid and dashed lines. }
	\label{model}
\end{figure}

We start from the minimum square lattice model, which is schematically plotted in Fig.~\ref{model}.  The unit cell includes two sublattices $A$ and $B$.  In momentum space, the Hamiltonian is given as \cite{F.Li, Y.X.Wang2012}:
\begin{align}
H_0(\boldsymbol k)=h_x\tau_x+h_y\tau_y+h_z\tau_z, 
\end{align}
where $h_x=-4t_1\text{cos}\varphi\text{cos}\frac{k_x}{\sqrt2}
\text{cos}\frac{k_y}{\sqrt2}$, $h_y=-4t_1\text{sin}\varphi\text{sin}\frac{k_x}{\sqrt2}
\text{sin}\frac{k_y}{\sqrt2}$, and $h_z=\Delta+2t_2[\text{cos}(\sqrt2 k_x)-\text{cos}(\sqrt2 k_y)]$.  The Pauli matrices $\tau$ acts on the sublattice degree of freedom.  $t_1$ and $t_2$ are the NN and next-nearest-neighbor (NNN) hopping integrals, respectively.  The nontrivial phase $\varphi$ is associated with the NN hopping and its sign depends on the direction of the bond.  $\Delta$ denotes the staggered sublattice potential.  Here we set the lattice constant as $a=1$.  Note $H_0$ is degenerate with spin and invariant under $SU(2)$ chiral rotation of the spin quantization axis, generated by $\sigma\otimes\tau_0$, with $\sigma$ acting on spin.
 
Around the Dirac points ${\boldsymbol K}=(\frac{\pi}{\sqrt2},0)$ and ${\boldsymbol K}'=(0,\frac{\pi}{\sqrt2})$, the low-energy Hamiltonian is expanded as:
\begin{align}
H_{\boldsymbol K}=2\sqrt2 t(\text{cos}\varphi q_x\tau_x-\text{sin}\varphi q_y\tau_y)+m_{\boldsymbol K}\tau_z, 
\end{align}
and
\begin{align}
H_{\boldsymbol K'}=2\sqrt2 t(\text{cos}\varphi q_y\tau_x-\text{sin}\varphi q_x\tau_y)+m_{\boldsymbol K'}\tau_z, 
\end{align} 
where $q_{x(y)}$ denotes the deviation of the wave vector from the Dirac point.  The fermion masses at the two Dirac points are given as $m_{\boldsymbol K}=\Delta-4t_1$ and $m_{\boldsymbol K'}=\Delta+4t_1$.  Note that the chiralities are opposite between the two Dirac points.  For simplicity, we choose $\varphi=\frac{\pi}{4}$, which leads to the isotropic Fermi velocity $v_F=2t_1$.  A tiny fluctuation of the phase from $\frac{\pi}{4}$ will not affect the topological property of the system.  We set $t_1=1$ as the unit of energy.   

In the presence of both on-site ($U$) and NN ($V$) repulsions, the extended short-range Hubbard Hamiltonian in real space is given as 
\begin{align}
H_I=U\sum_i n_{i\uparrow}n_{i\downarrow}
+V\sum_{\langle i,j\rangle,\sigma,\sigma'}n_{i\sigma}n_{j,\sigma'}.
\label{HI}
\end{align}
Here, $n_{i\sigma}=c_{i\sigma}^\dagger c_{i\sigma}$ is the fermionic number operator at site $i$ with spin $\sigma$, $c_{i\sigma}$ being the fermion annihilation operator. 

Besides the CI phase, the topological square lattice can support the existence of other phases, such as the two-dimensional Weyl semimetal and $2\pi-$flux topological semimetal.  There are also several works investigating the correlation effects on the different phases of topological square lattice.  For example, the Hubbard interaction on the $2\pi-$flux topological semimetal of square lattice was studied and the nematic phase was revealed~\cite{K.Sun}.  While the hard-core bosons with short-range interactions were considered on the square lattice~\cite{Y.F.Wang}, supporting the fractional quantum Hall states.

\section{Mean-field Method}

Because of the vanishing density of states around the Dirac points, any sufficiently weak local four-fermion interaction is an irrelevant perturbation in the sense of renormalization group so that the system is stable against weak interactions.  However, when the interactions increase to beyond the critical strength, various SSB phases may dominate the system.  The general symmetry-breaking order parameters are defined as \cite{B.Roy}:
\begin{align}
\Delta_{\mu\nu}=\langle \Psi^\dagger\sigma_\mu\otimes\tau_\nu\Psi\rangle, 
\end{align}
with the basis $\Psi=(c_{A\uparrow},c_{B\uparrow},c_{A\downarrow},c_{B\downarrow})^T$.  The corresponding phase transitions are continuous.  In fact, there may exist six types of long-range orders in the system, i.e., {\it bond density} $\Delta_{01}$, {\it current density} $\Delta_{02}$, {\it CDW} $\Delta_{03}$, {\it spin bond density} $\Delta_{j1}$, {\it spin current density} $\Delta_{j2}$ and {\it AFM} $\Delta_{j3}$, here $j=1,2,3$ representing the three spatial directions. 

To judge what kind of long-range orders are favored by interactions, we can get some insights from the normal state susceptibilities $\chi_{\mu\nu}$, as the critical strength of interaction is inversely proportional to $\chi_{\mu\nu}$.  After direct calculations, we obtain the static susceptibilities at zero external frequency and momentum as 
\begin{align}
\chi_{01}&=\chi_{j1}=\chi_{02}=\chi_{j2}
\nonumber\\
&=\frac{1}{2\pi v_F^2} 
(\frac{\Lambda^2}{\sqrt{\Lambda^2+m_{\boldsymbol K}^2}} +\frac{\Lambda^2}{\sqrt{\Lambda^2+m_{\boldsymbol K'}^2}}), 
\end{align}
and 
\begin{align}
\chi_{03}&=\chi_{j3}=\frac{1}{\pi v_F^2} (\frac{\Lambda^2+2m_{\boldsymbol K}^2} {\sqrt{\Lambda^2+m_{\boldsymbol K}^2}}
+\frac{\Lambda^2+2m_{\boldsymbol K'}^2} {\sqrt{\Lambda^2+m_{\boldsymbol K'}^2}}), 
\end{align} 
where $\Lambda$ is the ultraviolet cutoff of momentum.  The above results clearly show that the relation of $\chi_{03}>2\chi_{01}$ always holds, so the leading instabilities to interactions are the CDW $(Q=\Delta_{03})$ and AFM $(M_j=\Delta_{j3})$ orders.  The CDW order breaks the $C_4$ rotational symmetry in the system while the AFM order breaks the $SU(2)$ spin rotational symmetry.  

To find out the CDW and AFM orders explicitly, within the MF, we decouple the two-body interactions as:  
\begin{align}
n_{i\uparrow}n_{i\downarrow}
\simeq&\langle n_{i\uparrow} \rangle n_{i\downarrow} 
+\langle n_{i\downarrow}\rangle n_{i\uparrow} 
\nonumber\\
&-\langle c_{i\uparrow}^\dagger c_{i\downarrow}\rangle c_{i\downarrow}^\dagger c_{i\uparrow}
-\langle c_{i\downarrow}^\dagger c_{i\uparrow}\rangle c_{i\uparrow}^\dagger c_{i\downarrow}
+\text{const}, 
\end{align}
and 
\begin{align}
n_{i\sigma}n_{j\sigma'}\simeq\langle n_{i\sigma}\rangle n_{j\sigma'}+
\langle n_{j\sigma'}\rangle n_{i\sigma}+\text{const}. 
\end{align}
Here the terms in the brackets denote the fermion densities and spin densities averaged to the ground state and are solved self-consistently by diagonalizing the decoupled one-body Hamiltonian.  For the on-site interaction, we allow the existence of both inplane colinear and non-colinear terms.  In many previous works, the non-colinear terms are often neglected \cite{J.He1, J.He2, T.I.Vanhala, D.Prychynenko}.  But they can play important roles in forming the long-range inplane magnetic order \cite{A.M.Cook, W.Zheng, K.Jiang, V.S.Arun} and are kept here.  On the other hand, for the NN interaction, we only retain the fermion density terms, while other possible decoupling channels are dropped as they are related to such orders that have been demonstrated not to exist by the susceptibilities.  It should be noted that the constant terms must be included in calculating the total energy, as to determine the ground state of the system.

Since we are interested in the bulk as well as the edge physics, we perform calculations on a large-size lattice with periodic boundary condition and also on a ribbon structure.  We focus on the half-filling case, i.e., the average occupation numbers on the two sublattices satisfying the condition of $\langle n_{A\sigma}\rangle+\langle n_{B\sigma}\rangle=1$ for spin $\sigma$.  We can define the parameters $\Delta_\sigma$ to show the deviations of fermion occupation number on each sublattice from the average number $\frac{1}{2}$:
\begin{align}
\Delta_\sigma=\langle n_{A\bar\sigma}\rangle-\frac{1}{2}=\frac{1}{2}-\langle n_{B\bar\sigma}\rangle, 
\end{align}
with $\bar\sigma$ being opposite to $\sigma$.

For a given lattice size $L$, the number of fermion is $N=L$.  The Hatree-Fock (HF) solution will be the Slater determinant of spin states:
\begin{align}
|\psi\rangle=\prod_{n=1}^N \sum_{i=1}^L\sum_\sigma u_n(i,\sigma) c_{i,\sigma}^\dagger|0\rangle.  
\end{align}
The energy $\langle\psi|\hat H_0+\hat H_I|\psi\rangle$ will be minimized within the manifold of the Slater determinant.  From the wavefunction $|\psi\rangle$, any ground-state property of the model can be determined.  The HF approach allows the number of upspin and downspin particles to fluctuate and non-colinear spin orders to develop.  The self-consistent procedure may lead to a local minimum in energy.  To avoid this, we will take the random configurations as the initial trial states, to help the MF procedure locate the ground state corresponding to the global minimum in energy.

\section{Renormalized fermion mass and energy gap}

Here we consider when CDW or AFM order dominates the system, its effect on the band structures.  When only CDW order is present, the low-energy fermion masses are renormalized by interactions as:
\begin{align} 
&m_{\boldsymbol K\sigma}'=m_{\boldsymbol K} +U\Delta_\sigma-\frac{1}{2}zVQ, \label{mass1}
\\
&m_{\boldsymbol K'\sigma}'=m_{\boldsymbol K'} +U\Delta_\sigma-\frac{1}{2}zVQ.  \label{mass2}
\end{align}
here $z=4$ being the number of NN sites for square lattice.  From the above equations, we observe that the interaction-induced fermion mass renormalizations are the same for different Dirac points, but the roles of $U$ and $V$ are quite opposite in determining the fermion densities on the sublattices.  If the sublattice potential is positive $\Delta>0$, it can cause the density difference between the sublattices, $n_A<n_B$, which in turn leads to $\Delta_\sigma<0$ and $Q<0$.  This means that the onsite interactions will oppose the sublattice potential \cite{T.I.Vanhala} and the NN interactions will enhance the sublattice potential.  Similar arguments also hold for the negative sublattice potential $\Delta<0$.  Therefore we arrive at the conclusion that in the framework of MF, $U$ and $V$ will compete with each other, as the former tends to enlarge the density difference between the sublattices while the latter is to reduce the density difference. 

When the in-plane AFM order dominates the system, the self-consistent calculations show that the magnetizations are $m_{Ax}=-m_{Ay}=-m_{Bx}=m_{By}=m_\perp$ (see Fig.~\ref{mag_config}) and $m_{Az}=-m_{Bz}=m_z$.  The upspin and downspin bands are mixed and $\sigma_z$ is no longer a good quantum number.  However, the Dirac points remain at $\boldsymbol K$ and $\boldsymbol K'$.  The energy gap of the magnetic-ordered system at the Dirac points are calculated as: 
\begin{widetext}  
\begin{align}
&G_{\boldsymbol K1/2}=\sqrt 2\sqrt{(m'_{\boldsymbol K\uparrow})^2+(m'_{\boldsymbol K\downarrow})^2 +U^2m_\perp^2\mp(m'_{\boldsymbol K\uparrow}
+m'_{\boldsymbol K\downarrow})\sqrt{(m'_{\boldsymbol K\uparrow}-m'_{\boldsymbol K\downarrow})^2
+2U^2m_\perp^2}},   \label{gap1}
\\
&G_{\boldsymbol K'1/2}=\sqrt 2\sqrt{(m'_{\boldsymbol K'\uparrow})^2+(m'_{\boldsymbol K'\downarrow})^2
+U^2m_\perp^2\mp(m'_{\boldsymbol K'\uparrow}
+m'_{\boldsymbol K'\downarrow})\sqrt{(m'_{\boldsymbol K'\uparrow}-m'_{\boldsymbol K'\downarrow})^2 +2U^2m_\perp^2}}.   \label{gap2}
\end{align}
\end{widetext}
The subscript $1/2$ denotes the band with mixed spins.  Clearly, when $m_\perp=0$, the energy gaps given by the above equations are the same as those determined by Eqs.~(\ref{mass1}) and (\ref{mass2}). 

Here we find that both the possible CDW and AFM orders can change the energy gaps, but will not move the Dirac points.  This is different from our recent work \cite{Y.X.Wang2018} that the Chern insulator arises due to the two-dimensional spin-orbit coupling with Raman-assisted hoppings, in which Dirac points can be moved by interactions.

\section{Bulk phase diagrams}

\begin{figure*}
	\centering
	\includegraphics[width=16cm]{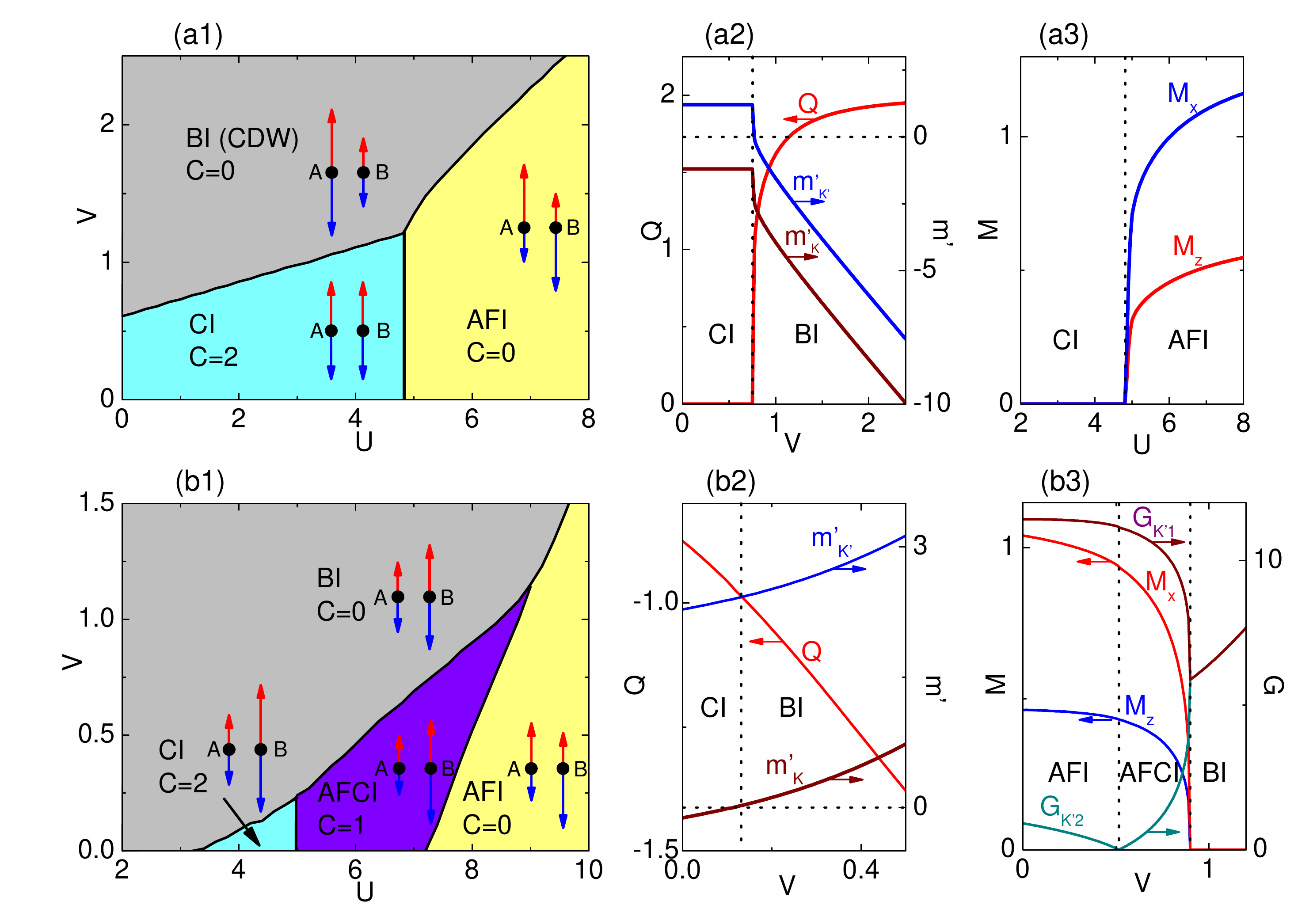}
	\caption{(Color online)  Phase diagrams in the parametric space $(U,V)$ of an interacting square lattice when the sublattice potential is vanishing as $\Delta=0$ (a1)-(a3) and nonvanishing as $\Delta=2$ (b1)-(b3).  The fermion occupation numbers are illustrated with arrows and the characteristic Chern numbers are also labeled.  (a2) The CDW order parameter $Q$ and the fermion masses $m_{K(K')}'$ vs $V$ when $U=1.2$.  (a3) The AFM order parameters $M_{x(z)}$ vs $U$ when $V=0.3$.  (b2) $Q$ and $m_{K(K')}'$ vs $V$ when $U=4.4$.  (b3) $M_{x(z)}$ and the energy gap  $G_{K'1(2)}$ vs V when $U=8$.  Note the double $y-$axis in (a2), (b2) and (b3).  The NNN hopping integral  $t_1=0.3$ and the number of unit cell is $N_x=N_y=32$. }
	\label{phase}
\end{figure*}

In this section, we calculate the interacting phase diagrams of a topological square lattice in the parametric space $(U,V)$.  The number of the unit cell in $x$ and $y$ directions is taken as $N_x=N_y=32$ and the periodic boundary conditions are used.  We have checked that the phase diagrams remain unchanged to a larger-size system.  Two cases are considered: the sublattice potential is vanishing as $\Delta=0$ in Figs.~\ref{phase}(a1)-(a3) and nonvanishing as $\Delta=2$ in Figs.~\ref{phase}(b1)-(b3).  The fermion occupation numbers of different spins on both sublattices are shown with arrows.  We also label the characteristic Chern number $C$, which is obtained by Fukui's algorithm \cite{T.Fukui}.  Indeed, the CI in Fig.~\ref{phase}(a1) and the BI in Fig.~\ref{phase}(b1) show certain robustness when $U$ and $V$ are weak and are protected by the energy gap.

First when $\Delta=0$, it shows that three distinct phases appear in Fig.~\ref{phase}(a1): the CI with $C=2$ (due to the degeneracy of two spin species), the BI and the antiferromagnetic insulator (AFI).  The distributions of fermions are equal on both sublattices as $\Delta=0$.  When $V$ increases to cross the critical strength, the CDW order appears.  In Fig.~\ref{phase}(a2) when $U=1.2$, the order parameter $Q$ is plotted vs $V$, where a clear phase transition can be seen at the critical $V_c^b=0.76$.  Meanwhile, Fig.~\ref{phase}(a2) also shows that the mass inversion occurs at the Dirac point $\boldsymbol{K}'$ along with the SSB and as a result, the system becomes topologically trivial as the BI.  On the other hand, increasing $U$ will drive the AFM order in $z-$direction as well as in $xy-$plane.  The corresponding order parameters $M_z$ and $M_x$ are plotted vs $U$ when $V=0.3$ in Fig.~\ref{phase}(a3), where the phase transition occurs at $U_c^b=4.9$.  As the symmetry-breaking phase is topologically trivial, it is called the AFI.  The two SSB critical lines merge at the tricritical point of $(U,V)=(4.86,1.35)$ in Fig.~\ref{phase}(a1), beyond which the AFI competes with CDW and the separating line is approximately linear as $V\sim 0.43U-0.78$.  Fig.~\ref{phase}(a1) clearly illustrates the mechanism of SSB driven by interactions on a topological square lattice.  Such a phase diagram is qualitatively similar with that of the interacting 3D line-node semimetal \cite{B.Roy} and hyperhoneycomb lattice \cite{S.W.Kim}.  

Next when $\Delta=2$, the inequivalent fermion numbers on the two sublattices are induced, giving rise to $n_A<n_B$ and $Q<0$.  The inequivalent fermion numbers will always exist as the interactions are increased, therefore the CDW order is ubiquitous in all phases.  In Fig.~\ref{phase}(b1), we observe that besides the three phases mentioned above, an additional phase of AFCI with $C=1$ appears, supporting a single gapless chiral edge mode.  It can be considered as interpolating between CI and AFI or between AFI and BI.  Being an interaction-driven phase, the AFCI spans the regimes of intermediate $U$ and low $V$ in the phase diagram and will be extended to stronger $U$ when $V$ increases.  To help judge the phase transitions, we plot two cases of $U=4.4$ and $U=8$ in Fig.~\ref{phase}(b2) and (b3), respectively.  In Fig.~\ref{phase}(b2), we can see that the CDW order parameter is negative $Q<0$ and the fermion mass inversion occurs at the Dirac point $\boldsymbol K$, resulting in the phase transition from CI to BI.  While in Fig.~\ref{phase}(b3), starting from the AFI phase, increasing $V$ can first make the gap $G_{\boldsymbol K'2}$ to be closed, so that band-2 becomes topologically nontrivial while band-1 remains topologically trivial.  As a result, the Chern number changes from $C=0$ to $C=1$ and the system enters the AFCI phase.  Further increasing $V$, the $SU(2)$ symmetry is restored and the system reenters the BI phase. 

\begin{figure}
	\includegraphics[width=7.2cm]{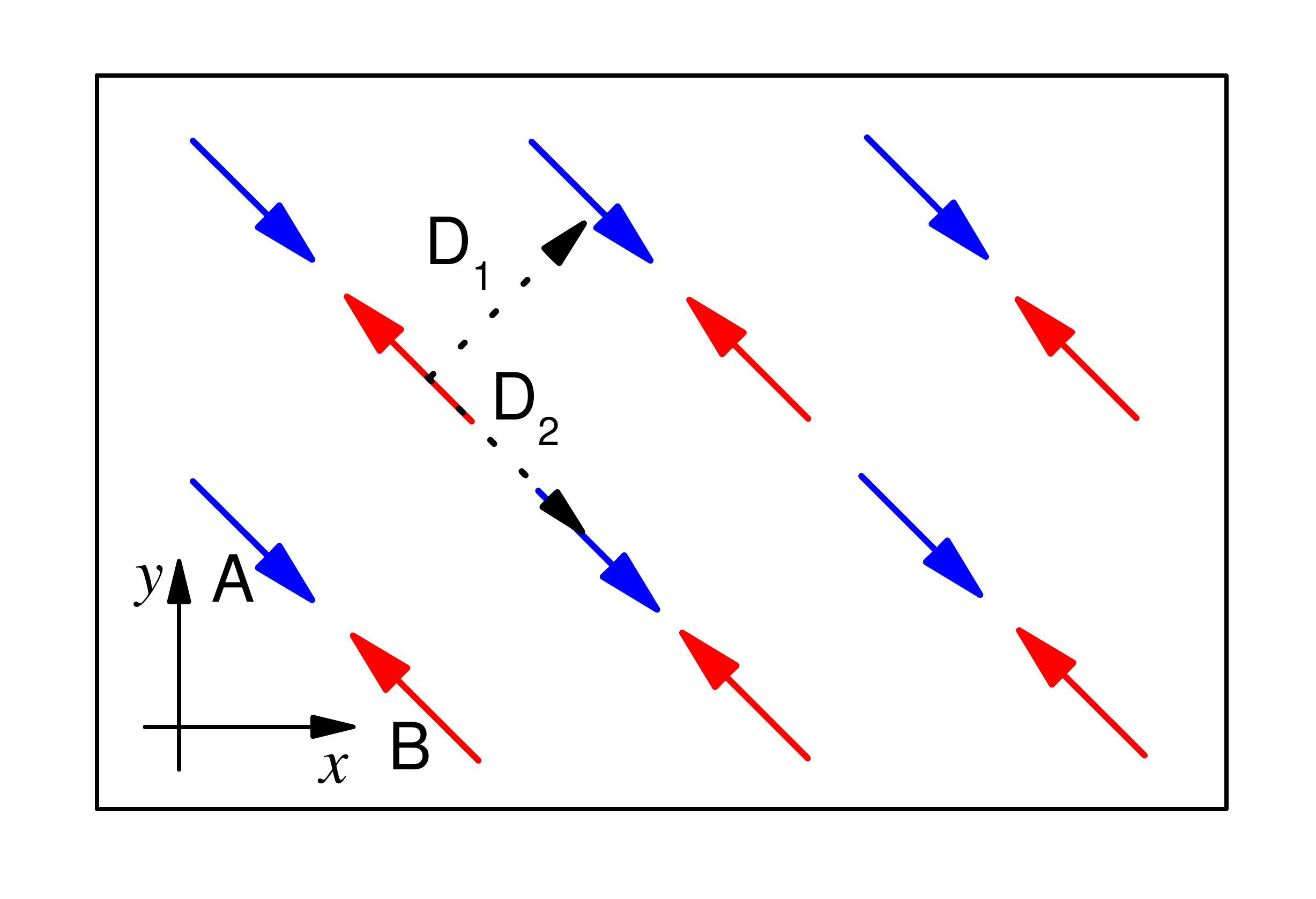}	
	\caption{(Color online) In-plane AFM configuration, exhibiting the checkerboard spin pattern with $m_{Ax}=-m_{Ay}=-m_{Bx}=m_{By}$.  ${\boldsymbol D}_1$ and ${\boldsymbol D}_2$ are the translational vectors by which all spins flip their signs. }
	\label{mag_config}
\end{figure}

Then why the nonvanishing sublattice potential can drive the AFCI phase and what is the underlying physical mechanism?  We can understand it from the viewpoint of breaking the TRS in magnetic topological insulators \cite{R.Mong,C.Fang,R.X.Zhang}.  The time-reversal operator is defined in the single-fermion sector of Hilbert space as 
\begin{align}
\Theta=K(i\sigma_y), 
\end{align}
with $K$ being the complex conjugate operator.  One can see that $\Theta$ is antiunitary and squares to minus the identity.  For the AFM order, it clearly breaks the TRS as the TRS can reverse all spins but leave the orbital and spatial components invariant, 
\begin{align}
\Theta H_M\Theta^{-1}=-H_M, 
\end{align}
where $H_M$ is the magnetic Hamiltonian and is derived from the decoupled interaction $H_I$ in Eq.~(\ref{HI}).  In Fig.~\ref{mag_config}, we plot the in-plane AFM configuration.  It shows that there exist some special lattice vectors, such as $\boldsymbol D_1=\frac{a}{\sqrt2}(1,1)$ and $\boldsymbol D_2=\frac{a}{\sqrt2}(1,-1)$, which, after translation, all spins will also reverse their directions:
\begin{align}
T_D H_M T_D^{-1}=-H_M,
\end{align}
with $T_D$ being the translational operator.  Then we define a new symmetry that combines the TRS and the translational symmetry as 
\begin{align}
\Theta_S=\Theta\otimes T_D, 
\end{align}
which is also antiunitary.  If the sublattice potential is vanishing, $\Delta=0$, we definitely have the commutation relation
\begin{align}
[H_M+H_0,\Theta_S]=0.  
\end{align}
That is, if we first do time-reversal operation to the system, and then make the translational operation, the electronic states will return to its original states.  This means that the TRS has not been truly broken so that the topologically nontrivial bands cannot appear.  While if the sublattice potential is explicitly nonvanishing,  $\Delta\neq0$, the above commutation relation does not hold anymore.  Then after the time-reversal operation, the electronic states cannot return to its original states under any spatial operation.  Therefore the TRS has been truly broken and the magnetic-ordered phase of AFCI with topologically nontrivial bands can appear. 

In a recent work about the interacting Kane-Mele model~\cite{K.Jiang}, the AFCI phase is demonstrated to occur in the 2D noncentrosymmetric system.  It should be emphasized that when the sublattice potential is nonvanishing, the inversion symmetry is indeed broken in honeycomb lattice, but is still preserved in square lattice.  So the nonvanishing sublattice potential is the necessary condition for AFCI.  It is also worth notable that in their work~\cite{K.Jiang}, the AFM order in $z$ direction cannot coexist with that in $xy$ plane, i.e., one magnetic order appears while another will be suppressed.  While in our work, the AFM orders always occur simultaneously in $z$ direction as well as in $xy$ plane.  This is because the topological square lattice keeps the spin-rotational $SU(2)$ symmetry and the corresponding static susceptibilities are equal $\chi_{13}=\chi_{33}$.  While for the interacting Kane-Mele model \cite{K.Jiang}, it includes spin-orbit coupling that breaks the spin-rotational $SU(2)$ symmetry.  As a result, the static susceptibilities are unequal in different directions, $\chi_{13}\neq\chi_{33}$, and either the AFM order in $xy-$plane or $z-$direction dominates the system, depending on the parameters. 

\begin{figure}
	\includegraphics[width=8.4cm]{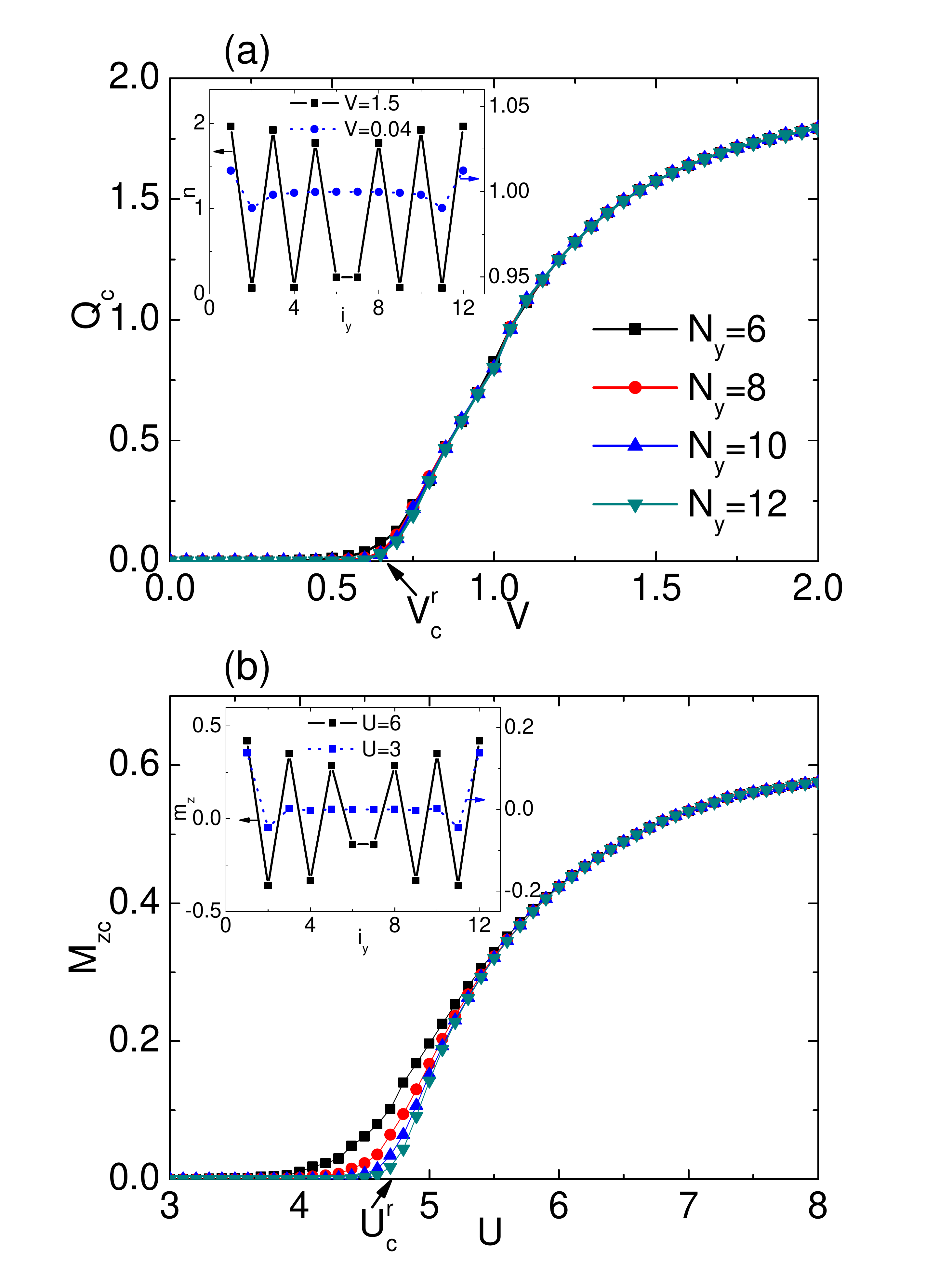}
	\caption{(Color online) Hatree-Fock self-consistent solution of the square lattice ribbon structure with $\Delta=0$, $t_1=0.3$ and $N_x=32$.  The plots are (a) the CDW order parameter of the central cell $Q_c$ vs $V$ with $U=1.2$, (b) the AFM order parameter $M_{zc}$ vs $U$ with $V=0.3$.  The critical interactions are shown to be $V_c^r\sim0.65$ and $U_c^r\sim4.7$.  The inset in (a) is the spatial variation of fermion density $n$ for different $V$ and (b) is the magnetization $m_z$ for different $U$, with $N_y=6$.  Note the double $y-$axis of the insets.  The legends are the same in both figures. }
	\label{ribbon}
\end{figure}

\section{Ribbon structure}

In this section, in order to see how the ribbon structure affects the SSB and the phase transitions, we investigate the interacting square lattice in the ribbon case.  As the experiments are performed on a finite system, here ribbon in an experimental sense means a much larger number of sites along one direction compared with that along another direction, \textit{e.g.} in Refs.~\cite{M.Mancini, B.K.Stuhl}.  We take the periodic boundary condition in $x-$direction, but the open boundary condition in $y-$direction for the ribbon system.  

With the parameters being set to be the same as Fig.~\ref{phase}(a1), the interacting square lattice ribbon structure is solved in the MF level.  To reduce the effect of boundary as much as possible, we choose the CDW and AFM order parameters as: 
\begin{align}
Q_c&=\big(\langle n_{A\uparrow}^c\rangle
+\langle n_{A\downarrow}^c\rangle\big)
-\big(\langle n_{B\uparrow}^c\rangle
+\langle n_{B\downarrow}^c\rangle\big), 
\end{align}
and
\begin{align}
M_{zc}&=\big(\langle n_{A\uparrow}^c\rangle
-\langle n_{A\downarrow}^c\rangle\big)
-\big(\langle n_{B\uparrow}^c\rangle
-\langle n_{B\downarrow}^c\rangle\big), 
\end{align}
with $n_{\alpha\sigma}^c$ being the local fermion number operator for sublattice $\alpha$ and spin $\sigma$ in the central cell.  The numerical results are plotted in Fig.~\ref{ribbon}, where $Q_c$ in Fig.~\ref{ribbon}(a) are in good consistent to different size $N_y$ while $M_{zc}$ in Fig.~\ref{ribbon}(b) exhibit quick convergence to larger size.  These suggest that our results are reliable in the thermodynamic limit.  In Fig.~\ref{ribbon}(a), the CDW order happens when $V>V_c^r\sim0.65$ and then $Q_c$ gradually tends to the saturation value.  While in Fig.~\ref{ribbon}(b), the AFM order occurs when $U>U_c^r\sim4.7$.  Compared with the critical interactions in the bulk system with the same parameters, $V_c^b=0.76$ in Fig.~\ref{phase}(a2) and $U_c^b=4.9$ in Fig.~\ref{phase}(a3), the ribbon structure exhibits weaker critical values.  

\begin{figure}
	\includegraphics[width=8.8cm]{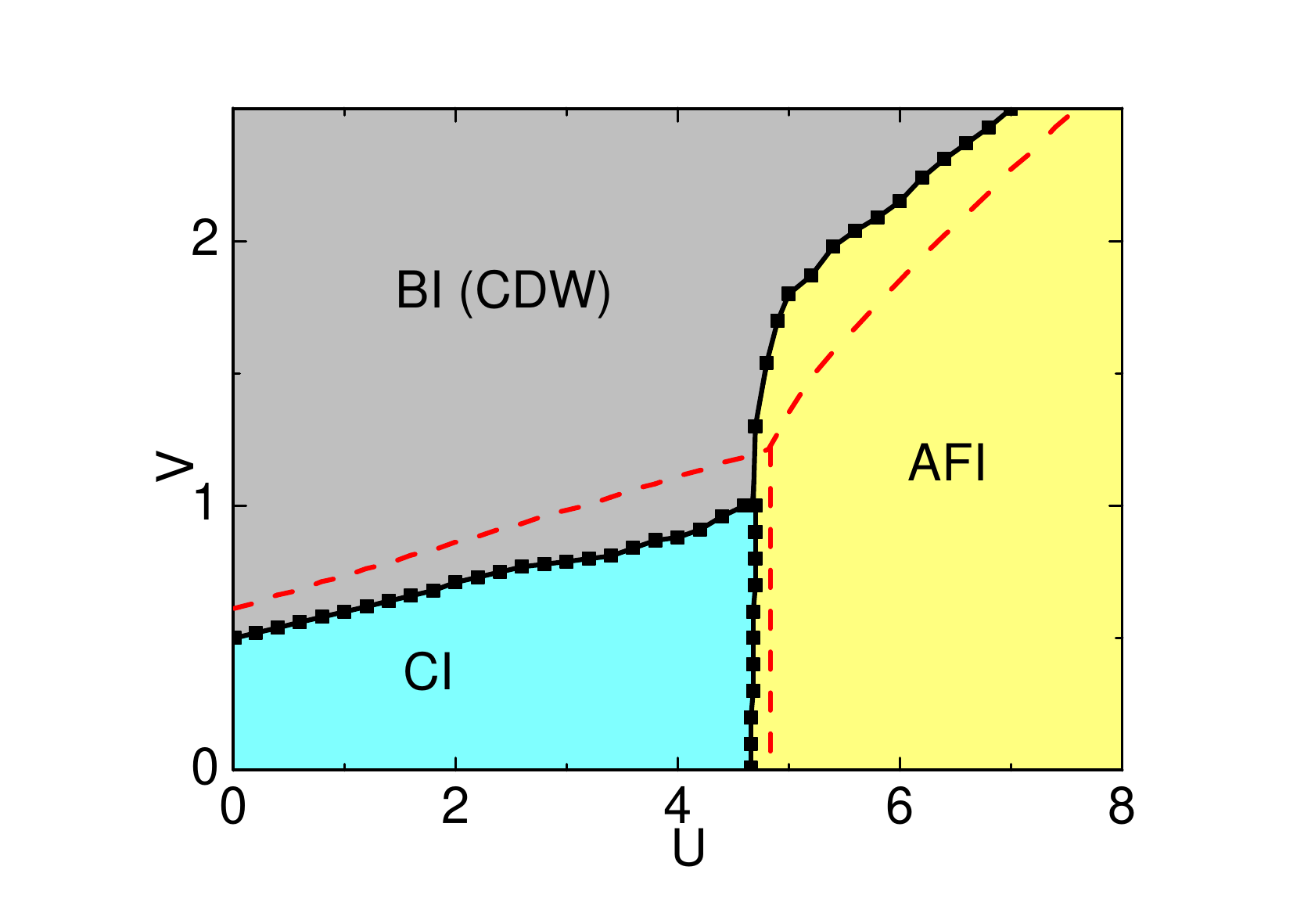}
	\caption{(Color online) The phase diagram obtained from the interacting square lattice ribbon structure with $N_x=32$, $N_y=12$, $\Delta=0$ and $t_1=0.3$.  The dashed line are the bulk phase boundaries from Fig.~\ref{phase}(a1) for comparison.}
	\label{Fig5}
\end{figure}

The whole phase diagram calculated from the ribbon structure for $\Delta=0$ is plotted in Fig.~\ref{Fig5}.  It shows that in the ribbon phase diagram,  when compared with the bulk one, whose phase boundaries have also been plotted in Fig.~\ref{Fig5} by the red dashed lines, two aspects are worth notable:  (i) for the transitions from CI to BI and CI to AFI, the phase boundary is shifted to lower $V$ and lower $U$, respectively; (ii) for the transition from AFI to BI, the phase boundary is pushed to higher $V$.  In the work by Cao and \textit{et.al} \cite{J.Cao}, the authors obtain the phase boundaries in the ribbon structure by counting the edge states in the gap.  They point out when compared with the bulk phase diagram, the phase boundaries of the ribbon are greatly modulated and even the structure of the phase diagram is dramatically changed.  Here in Fig.~\ref{Fig5}, we find that the structure of the ribbon phase diagram is kept unchanged, but only the phase boundaries shift.  

The shiftness of the phase boundaries can be explained as follows.  When the system lies in the CI phase, the gapless edge states are supported in the ribbon structure, leading to the finite density of states around the Dirac points.  Consequently, the edge sites undergo weak interaction instabilities before the bulk acquires any orderings.  In the insets of Figs.~\ref{ribbon}(a) and (b), we can see that even when the interactions are weak and below the critical values, the local CDW and AFM order at the edge sites are still induced by interactions, but will quickly vanish at the neighboring sites.  Then with the increasing of interactions and through the proximity effect, these local instabilities in turn give rise to the long-range orders in the whole system, which happens at weaker interactions than the bulk system.  This explains the transition from CI to BI as well as CI to AFI.  While for the transition from AFI to BI, it can be ascribed to the fact that, in the ribbon structure, the number of degrees of freedom participating in the onsite interaction $U$ remains unchanged, while that participating in the NN interaction $V$ is reduced due to the existence of boundary.  As $U$ is the main factor leading to the AFM order and $V$ to the CDW order, thus to drive the system from AFI into BI, a stronger $V$ is needed when compared with the bulk system, as shown in Fig.~\ref{Fig5}.  That is, the interaction-driven AFM order is more stable than CDW in the ribbon structure when they compete with each other.

\section{Discussions and Summaries}

Experimentally, there have been no reports about the AFM ordered TI in real electronic materials so far.  It is believed that in the CuO$_2$ layer of any parent compound of cuprate superconductors \cite{C.C.Tsuei} as well as the vacancy-doped iron-based superconductor \cite{M.Wang}, the ground states can exhibit the AFM spin structures, but their energy bands do not own any topological properties.  We hope that the interaction-driven AFCI phase can be demonstrated in cold-atom system, where all physical factors, including the interactions, can be controlled precisely. In particular, the sublattice potential can be modulated by creating an energy offset between neighboring sites \cite{G.Jotzu, L.Tarruell}. The topologically nontrivial bands can be detected by measuring the orthogonal drift of atoms after applying a constant force \cite{G.Jotzu, M.Aidelsburger2} while the long-range AFM order of atoms can be measured from the Bragg scattering of light \cite{A.Mazurenko, T.A.Corcovilos}.    

To summary, in this work we have studied the extended Hubbard model on a topological square lattice that supports the CI.  We reveal that the effects of interactions incorporate changing the energy gap as well as inducing the SSB.  The correlated AFCI phase is demonstrated to exist only when the TRS is truly broken by the explicitly nonvanishing sublattice potential.  We also consider the interacting ribbon structure and find the influence of finite width of the system on the phase diagram.  Although the MF approach provides an initial understanding about the correlation effect in CI, we suggest the qualitative properties of the interacting phase diagram can be retained when highly-advanced techniques, such as the dynamical MF method \cite{T.I.Vanhala} or dynamical cluster approximation \cite{J.Imriska}, are applied.

\section{Acknowledgments}

We would like to thank Fuxiang Li and Biao Huang for many helpful discussions.  This work was supported by NSFC (Grant No. 11804122) and China Scholarship Council (No. 201706795026).

\end{document}